\documentclass[aps,twocolumn,showpacs, preprintnumbers,amsmath,amssymb]{revtex4}

\usepackage{graphicx}
\usepackage{amsmath}

\renewcommand{\rho}{\varrho}
\renewcommand{\phi}{\varphi}
\renewcommand{\vec}{\mathbf}

\begin{document}

\title{Variational average-atom model of electron-ion plasma with correlations and quantum bound electrons}

\author{R. Piron$^1$\footnote{Corresponding author\\Electronic address: robin.piron@cea.fr} and T. Blenski$^2$}
\affiliation{$^1$CEA, DAM, DIF, F-91297 Arpajon, France.}
\affiliation{$^2$Laboratoire ``Interactions, Dynamiques et Lasers'', UMR 9222, CEA-CNRS-Universit\'e Paris-Saclay, Centre d'\'Etudes de Saclay, F-91191 Gif-sur-Yvette Cedex, France.}

\date{\today}

\begin{abstract}
In the present paper, we propose a variational average-atom model of electron-ion plasma performing a quantum treatment of bound electrons and accounting for correlations (VAAQBEC). This model addresses the correlation functions in a weakly-coupled plasma, while also accounting self-consistently for the ion average shell structure. This is done at the price of treating the free electrons classically, whereas bound electrons are treated quantum-mechanically.
When ions are approximated by point-like particles, the present approach yields the usual Debye-H\"{u}ckel corrections to the orbital energies and chemical potential. If one disregards the interactions of continuum electrons, the present approach yields ion-ion correlation corrections through a self-consistent one-component-classical-plasma contribution.
Comparisons are presented with the broadly-used continuum-lowering approach of Stewart and Pyatt \cite{StewartPyatt66} and with the dense-plasma average-atom models INFERNO \cite{Liberman79,Liberman82} and VAAQP \cite{Blenski07b,Piron11}, on warm silicon and hot iron cases.
\end{abstract}


\maketitle

\section{Introduction}

The typical model of ideal plasma in thermal equilibrium is the Saha model \cite{Saha20,Saha21}. In this model, the plasma is viewed as an ideal-gas mixture, where the species are the various ion electron states, plus the free electrons. The shell structures of the ion species are fixed, and calculated using a separate model, which can be of various degree of approximation (screened-hydrogenic, quantum detailed configuration accounting, quantum detailed level accounting...). An average-atom equivalent \cite{Mayer47} to the Saha model exists, with several possible approximations for the calculation of the average shell structure (see, for instance, \cite{Mancini85}).

In these ideal plasma models, the ion shell structure is impacted neither by the effect of the interaction between ions, nor by the effect of the polarization of continuum electrons around the ions. Such effects, whose relevance grows with density, are often designated under the generic name of ``density effects'' on the shell structure. 

In order to account for these effects, corrections to the isolated-ion orbital energies were developped \cite{EckerWeizel56,EckerKroll63,StewartPyatt66}.  In parallel, some authors have tried to extend the Thomas-Fermi ion-in-cell model \cite{Feynman49} to the quantum description of electrons. This led to the development of dense-plasma models such as Rozsnyai's model \cite{Rozsnyai72}, INFERNO \cite{Liberman79,Liberman82} or VAAQP \cite{Blenski07b,Piron11}, in which these ``density effects'' are built in the model, to some degree of approximation.

In Rozsnyai's model, polarization of the continuum electrons is accounted for using a semi-classical (i.e. Thomas-Fermi) model of the continuum electrons, whereas bound electrons are treated through a band model, as in solid-state physics. The effects of the neighboring ions are accounted for through the neutrality of the Wigner-Seitz sphere. Due to this neutrality, the self-consistent potential goes to zero at the Wigner-Seitz radius.

In the INFERNO model, polarization of the continuum electrons is accounted for using the same quantum formalism as for bound electrons. The effects of the neighboring ions are treated in a way similar to that of Rozsnyai's model. The Wigner-Seitz sphere is neutral, and the self-consistent potential is zero outside the sphere.

In the VAAQP model, polarization of the continuum electrons is also accounted for using the same formalism as for bound electrons. There is no restriction of the potential range to the Wigner-Seitz sphere. Instead, the effects of the non-central ions are accounted for through the interaction of all electrons with a non-central-ion charge density, which is assumed to have the form of a Heaviside function. 

In all these dense-plasma models, it is assumed that the non-central ions form a statistical cavity around the central ion. That is, non-central ions have zero probability to enter the Wigner-Seitz sphere, and are uniformly distributed outside the Wigner-Seitz sphere. This cavity assumption comes from the qualitative behavior of the correlation function in strongly-coupled one-component classical fluids with strong repulsive interaction at short distances.

In order to further improve over these models, there is an increasing interest for including a self-consistent accounting for ion-ion correlations in average-atom models \cite{Starrett12, Chihara16}. 
Among the deepest issues which faces such a theoretical effort is the lack of practical formulation of the quantum theory of correlation functions in fluids (see, for instance \cite{Percus05}).

On the other hand, there has always been an interest for the extension of dense-plasma models towards the low-coupling regime. From a practical point of view, dense plasma models such as INFERNO or VAAQP can become computationally expensive in the case of low-density plasmas. Moreover, it is expected that, in the low-coupling regime, the ion-ion correlation function departs strongly from the cavity shape, going smoothly to the ideal-gas form. This was, for instance, among the motivation for Crowley's work \cite{Crowley90}. 

In the present paper, we propose a Variational Average-Atom model of electron-ion plasma performing a Quantum treatment of Bound Electrons and accounting for Correlations  (VAAQBEC). This model addresses the three 2-particle correlation functions in a weakly-coupled plasma, while also accounting self-consistently for the ion average shell structure. In the present formulation, this is done at the price of treating the free-electrons classically, whereas bound electrons are treated quantum-mechanically. Such a splitting between bound and continuum electrons may break the continuity of observables and is only justified for weak polarization of the continuum electrons.

Unlike the broadly-used continuum-lowering models, the present model naturally introduces the screening of the self-consistent potential. It thus leads rigorously to a finite number of bound orbitals, without resorting to an \emph{ad-hoc} suppression of bound states after shifting the energies. It also accounts for the effects of the screened potential on the radial wave-functions, which is known to have an impact on the oscillator strengths (see, for instance, \cite{Shore75}) and more generally on all atomic cross-sections. However, since this model relies on a classical treatment of the continuum electrons, it notably disregards the resonances in the continuum.

In Sec.~\ref{sec_AAII}, we present an average-atom model of isolated ion. This kind of model is not new but the present, variational, derivation is used to introduce the perspective and notations for the following. In Sec.~\ref{sec_DHAAM}, we derive the equations for the VAAQBEC model. In Sec.~\ref{sec_DHAAM_CL}, we address a version of the model in which ions are considered as point-like, and show that this leads to the usual continuum-lowering picture. In Sec.~\ref{sec_DHAAM_OCP}, we focus on another version of the model, in which free-electron interactions are neglected. The latter approximation leads to a one-component-plasma approach to the correlations. Finally, we discuss in Sec.~\ref{sec_Results} some results from the VAAQBEC approach and give some comparisons with results from the Stewart-Pyatt continuum-lowering approach \cite{StewartPyatt66} as well as from the INFERNO and VAAQP dense-plasma models.

\section{Average-atom model of isolated ion from a variational perspective\label{sec_AAII}}

In the Saha model, a strong distinction is made between ``bound electrons'', which participate in the ion shell structure, and ``free electrons'', which constitute a particular species of the ideal-gas mixture. The equations of the Saha equilibrium model are obtained through a minimization of the free energy of the system with respect to the species populations, while also requiring the neutrality of the plasma. Then, only the populations, and not the quantities related to the shell structure of the ions stem from the model.

In order to obtain an average-atom ``equivalent'' of the Saha model, let us consider a system composed of:
\begin{itemize}
\item non-interacting ions, each ion having finite charge and the same, average, shell structure. Each ion is thus an interacting system of a nucleus of charge $Z$ and $N_\text{b}\leq Z$ bound electrons, 
\item a uniform, ideal electron gas gathering all electrons that belong to the continuum.
\end{itemize}
The free energy per ion of such a system can be written as follows:
\begin{align}
F&\left\{\left\{f_\alpha\right\},\underline{v}_\text{trial},n_0;n_\text{i},T\right\}\nonumber\\
&=F_0^\text{i}(n_\text{i},T)+F_0^\text{e}(n_0;n_\text{i},T)+\Delta F_1\left\{\left\{f_\alpha\right\},\underline{v}_\text{trial}\right\}
\label{eq_AAII_free_energy}
\end{align}
where the functional dependencies are underlined.
Here, the $F_0^\text{i}$ term corresponds to the contribution of the nuclei ideal-gas and $F_0^\text{e}$ corresponds to the contribution of the free-electron ideal gas:
\begin{align}
F_0^\text{i}=\frac{1}{n_\text{i}}\left(\ln\left(n_\text{i}\Lambda_\text{i}(T)^3\right)-1\right)\ ; \ F_0^\text{e}=\frac{f_0(n_0,T)}{n_\text{i}}
\end{align}
where $\Lambda_\text{i}$ is the nucleus thermal length, and $f_0$ is the free energy per unit volume of an electron ideal gas of density $n_0$. This electron gas can be considered as degenerate or not.

The $\Delta F_1$ term corresponds to the free-energy of the average ion shell structure, that is: the interacting system of bound electrons and the nucleus. We treat this system in a density-functional-like formalism, that is, we split $\Delta F_1$ into the three contributions:
\begin{align}
\Delta F_1\left\{\left\{f_\alpha\right\}, \underline{v}_\text{trial};T\right\}
=\Delta F_1^0 + \Delta F_1^\text{el} + \Delta F_1^\text{xc}
\end{align}
where $\Delta F_1^0$ is the kinetic-entropic contribution to the free energy, $\Delta F_1^\text{el}$ is the direct electrostatic contribution, and $\Delta F_1^\text{xc}$ is the exchange-correlation contribution.
\begin{align}
\Delta F_1^0&\left\{\left\{f_\alpha\right\}, \underline{v}_\text{trial};T\right\}
\nonumber\\
&=\sum_\alpha \left(
f_\alpha\left(\varepsilon_\alpha-\int d^3r \left\{v_\text{trial}(r)|\varphi_\alpha(\vec{r})|^2\right\}\right)
-T s_\alpha\right)\\
&=\sum_\alpha \left(
f_\alpha\varepsilon_\alpha-\int d^3r \left\{n(r)v_\text{trial}(r)\right\}
-T s_\alpha\right)\label{eq_AAII_dF10}
\end{align}
where the $\alpha$-indices label the non-degenerate 1-electron states (i.e. $n,\ell,m_\ell,m_s$ states in the non-relativistic case, $n,\ell,j,m_j$ states in the relativistic case). $f_\alpha$ is the mean occupation number of the 1-electron state $\alpha$, and the corresponding contribution to the entropy of the effective non-interacting system is:
\begin{align}
s_\alpha=s(f_\alpha)=-k_\text{B}\left(f_\alpha\ln\left(f_\alpha\right)
+(1-f_\alpha)\ln\left(1-f_\alpha\right)\right)
\label{eq_dft_entropy}
\end{align}
$\varepsilon_\alpha$ and $\varphi_\alpha(\vec{r})$ are shorthand notations for $\varepsilon_\alpha\left\{\underline{v}_\text{trial}\right\}$ and $\varphi_\alpha\left\{\underline{v}_\text{trial};\vec{r}\right\}$, respectively. These are the eigenvalues and wave-functions of the 1-electron states obtained in the trial potential $v_\text{trial}$. In the non-relativistic case, they are obtained solving the 1-electron Schr\"{o}dinger equation:
\begin{align}
-\frac{\nabla^2}{2}\varphi_\alpha(\vec{r})+(v_\text{trial}(r)-\varepsilon_\alpha)\varphi_\alpha(\vec{r})=0
\end{align}
We take the convention of normalizing the $\varphi_\alpha$ to unity.
$n(r)$ is a shorthand notation for $n\left\{\left\{f_\alpha\right\},\underline{v}_\text{trial};r\right\}$, the electron density of the effective independent-particle system defined by the trial potential $v_\text{trial}$ and the occupation numbers $\left\{f_\alpha\right\}$:
\begin{align}
n(r)=\sum_\alpha f_\alpha|\varphi_\alpha(\vec{r})|^2\label{eq_AAII_n}
\end{align}

The direct electrostatic contribution can be written as:
\begin{align}
\Delta F_1^\text{el}&\left\{\left\{f_\alpha\right\}, \underline{v}_\text{trial}\right\}
=\tilde{\Delta F}_1^\text{el}\left\{\underline{n}\right\}\nonumber\\
&=e^2\int d^3r\left\{\frac{-Z n(r)}{r}\right\}
+\frac{e^2}{2}\int d^3r d^3r'\left\{\frac{n(r)n(r')}{|\vec{r}-\vec{r}'|}\right\}
\label{eq_AAII_dF1el}
\end{align}
The exchange-correlation contribution can be approximated by:
\begin{align}
\Delta F_1^\text{xc}\left\{\left\{f_\alpha\right\}, \underline{v}_\text{trial};T\right\}
&=\tilde{\Delta F}_1^\text{xc}\left\{\underline{n};T\right\}\nonumber\\
&=\int d^3r\left\{f_\text{xc}\left(n(r),T\right)\right\}\label{eq_AAII_dF1xc}
\end{align}
where $f_\text{xc}$ is the local exchange-correlation free energy per unit volume.

In order to obtain the equations of the model, we minimize the free energy per ion, requiring the additional constrain of overall neutrality:
\begin{align}
Z - \sum_\alpha f_\alpha =\frac{n_0}{n_\text{i}}
\label{eq_neutr_AAII}
\end{align}
Consequently, we define the following functional:
\begin{align}
\Omega\left\{\left\{f_\alpha\right\},\underline{v}_\text{trial},n_0;n_\text{i},T\right\}\equiv F+\mu\left(Z-\sum_\alpha f_\alpha-\frac{n_0}{n_\text{i}}\right)
\end{align}
Performing the unconstrained minimization of $\Omega$:
\begin{align}
\frac{\delta \Omega}{\delta v_\text{trial}(r)}=\frac{\partial \Omega}{\partial f_\alpha}=\frac{\partial \Omega}{\partial n_0}=0
\end{align}
we get the equations of the average-atom isolated-ion (AAII) model:
\begin{align}
&v_\text{trial}(r)=v_\text{el}(r)+v_\text{xc}\left(n(r)\right)\label{eq_AAII_vtrial}\\
&f_\alpha = \frac{1}{e^{\beta(\varepsilon_\alpha-\mu)}+1}\label{eq_AAII_FermiDirac}\\
&\mu=\mu_0(n_0,T)\label{eq_AAII_mu}
\end{align}
where $v_\text{el}(r)$ is a shorthand notation for $v_\text{el}\left\{\underline{n};r\right\}$, defined as follows:
\begin{align}
v_\text{el}\left\{\underline{n};r\right\}
=\frac{\delta \tilde{\Delta F}_1^\text{el}}{\delta n(r)}=-\frac{Ze^2}{r}+e^2\int d^3r'\left\{\frac{n(r')}{|\vec{r}-\vec{r}'|}\right\}
\label{eq_def_vel}
\end{align}

We stress that in this model, as in the Saha model, there is a strong distinction between bound and free electrons, since any electron that belongs to the continuum is considered as non-interacting, whereas bound electrons interact both with the other bound electrons of the same ion and with its nucleus, through the self-consistent potential $v_\text{trial}$. 

We can also derive the electron pressure that stems from the AAII model:
\begin{align}
P&=n_\text{i}^2\frac{\partial F_\text{eq}}{\partial n_\text{i}}
=n_\text{i}^2\left.\frac{\partial \Omega}{\partial n_\text{i}}\right|_\text{eq}\\
&=n_\text{i}k_\text{B} T-f_0(n_0,T)+n_0\mu_0(n_0,T)
\label{eq_pressure}
\end{align}
This corresponds to the pressure of the ideal-gas mixture of ions and free-electrons. 


\begin{figure}[t]
\centerline{\includegraphics[width=8cm]{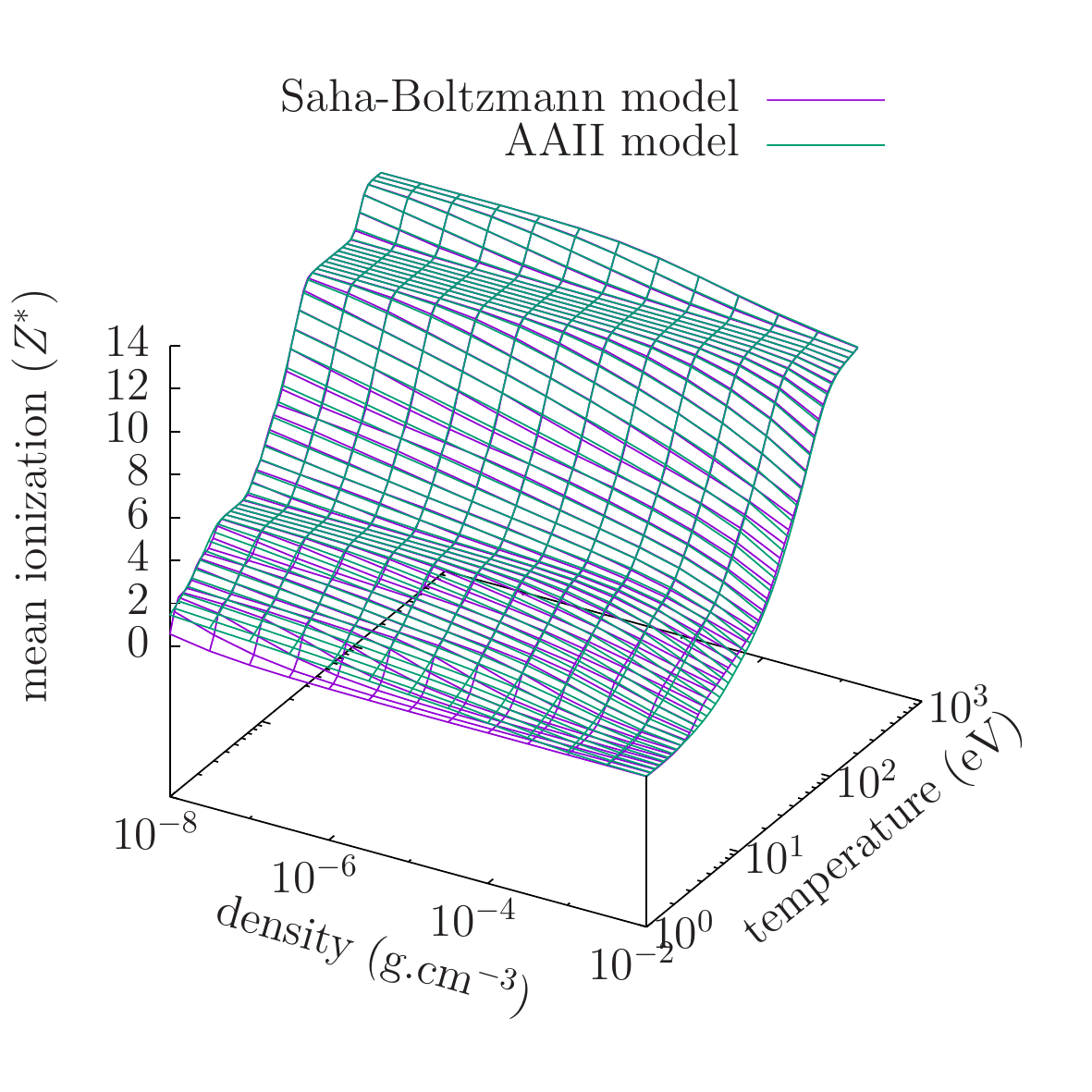}}
\caption{Mean ionization of silicon stemming from the Saha equilibrium model, with detailed configuration accounting of the ion electron states, and from the average-atom model of isolated ion (AAII).
\label{fig_Saha_AAII}}
\end{figure}

Fig.~\ref{fig_Saha_AAII} presents a comparison between the mean ionization: $Z^*\equiv n_0/n_\text{i}$ obtained from the Saha equilibrium model, using a detailed configuration accounting for the ion electron states, and the present AAII model, for the case of silicon. Though the results differ slightly, they are rather close. The qualitative behavior of decreasing mean ionization when density increases is similar, exhibiting clearly the absence of pressure ionization in these models.


Finally, let us mention that the expressions of Eqs~\eqref{eq_AAII_dF10}, \eqref{eq_AAII_dF1el}, \eqref{eq_AAII_dF1xc} for the free-energy, with the density of Eq.~\eqref{eq_AAII_n}, correspond to the free energy expression of the VAAQP model \cite{Blenski07b,Piron11}, when neglecting non-central ions and all interactions of the continuum electrons~\footnote{In the references \cite{Blenski07b,Piron11}, the trivial contribution $F_0^\text{i}$ of the ion ideal gas is disregarded, and the convention on the sign of $v_\text{el}$ is opposite.}. The former two assumptions are consistent with the hypotheses of the Saha model: zero coupling of the mixture of ions and free electrons. Thus, the VAAQP model recovers, in the Saha limit, the present AAII model. One can notably check that the pressure formula of Eq.~\eqref{eq_pressure} is identical to the pressure formula of the VAAQP model when the Wigner-Seitz radius tends to infinity. However, let us stress that, even if the cavity hypothesis of the VAAQP model plays no role in the zero-coupling Saha limit, it may be strongly irrelevant in the low-coupling regime.

\section{Derivation of the VAAQBEC model \label{sec_DHAAM}}
Let us now consider an interacting system composed of $N_\text{i}$ ions, whose quantum bound electrons form the density $n_\text{b}(\vec{r})$, and $N_\text{f}$ classical free electrons, in a volume V large compared to the shell structure of the ions.
The direct Coulomb interaction energy of such a system can be written as:
\begin{align}
U_\text{el}&\left\{N_\text{i},N_\text{f};\vec{R}_1...\vec{R}_{N_\text{i}};\vec{r}_1...\vec{r}_{N_\text{f}};\underline{n}_\text{b};V\right\}\nonumber\\
=&\frac{e^2}{2}\sum_{i=1}^{N_\text{i}}\sum_{j=1}^{N_\text{i}}\frac{Z^2}{|\vec{R}_i-\vec{R}_j|}
+\frac{e^2}{2}\sum_{i=1}^{N_\text{f}}\sum_{j=1}^{N_\text{f}}\frac{1}{|\vec{r}_i-\vec{r}_j|}
\nonumber\\
&+\frac{e^2}{2}\int_V d^3rd^3r'\left\{\frac{n_\text{b}(\vec{r})n_\text{b}(\vec{r}')}{|\vec{r}-\vec{r}'|}\right\}
+e^2\sum_{i=1}^{N_\text{i}}\sum_{j=1}^{N_\text{f}}\frac{-Z}{|\vec{R}_i-\vec{r}_j|}
\nonumber\\
&+e^2\sum_{i=1}^{N_\text{i}}\int_V d^3r \left\{\frac{-Zn_\text{b}(\vec{r})}{|\vec{R}_i-\vec{r}|}\right\}
+e^2\sum_{i=1}^{N_\text{f}}\int_V d^3r \left\{\frac{n_\text{b}(\vec{r})}{|\vec{r}_i-\vec{r}|}\right\}
\end{align}
The neutrality condition for this system can be written as:
\begin{align}
N_\text{i}\,Z=N_\text{f}+\int_V d^3r\left\{n_\text{b}(\vec{r})\right\}
\end{align}
Since we are constructing an average-atom model, we approximate the bound-electron density $n_\text{b}(\vec{r})$ by a superposition of identical, spherically-symmetric contributions $n(r)$, each corresponding to the shell structure of an ion:
\begin{align}
n_\text{b}(\vec{r})=\sum_{i=1}^{N_\text{i}}n(|\vec{r}-\vec{R}_i|)
\label{eq_one_center_approx}
\end{align}
The direct Coulomb interaction energy can then be readily rewritten as:
\begin{align}
U_\text{el}&\left\{N_\text{i},N_\text{f};\vec{R}_1...\vec{R}_{N_\text{i}};\vec{r}_1...\vec{r}_{N_\text{f}};\underline{n};V\right\}\nonumber\\
=&\frac{1}{2}\sum_{i=1}^{N_\text{i}}\sum_{\substack{j=1\\i\neq j}}^{N_\text{i}}
v_\text{ii}\left\{\underline{n};|\vec{R}_i-\vec{R}_j|\right\}+\frac{1}{2}\sum_{i=1}^{N_\text{f}}\sum_{\substack{j=1\\i\neq j}}^{N_\text{f}}
v_\text{ee}(|\vec{r}_i-\vec{r}_j|)
\nonumber\\
&+\sum_{i=1}^{N_\text{i}}\sum_{j=1}^{N_\text{f}}
v_\text{ie}\left\{\underline{n};|\vec{R}_i-\vec{r}_j|\right\}
+N_\text{i}\tilde{\Delta F}_1^\text{el}\left\{\underline{n}\right\}
\end{align}
where we have defined:
\begin{align}
v_\text{ii}\left\{\underline{n};V;R\right\}\equiv&\frac{Z^2e^2}{R}
+ e^2\int_V d^3r d^3r'\left\{\frac{n(r)n(r')}{|\vec{r}-\vec{r}'+\vec{R}|}\right\}
\nonumber\\
&-2e^2 \int_V d^3r \left\{\frac{Zn(r)}{|\vec{r}-\vec{R}|}\right\}
\\
v_\text{ie}\left\{\underline{n};V;r\right\}\equiv&
\frac{-Ze^2}{r}+e^2\int_V d^3r' \left\{\frac{n(r')}{|\vec{r}-\vec{r}'|}\right\}\\
v_\text{ee}(r)\equiv&\frac{e^2}{r}
\end{align}
and where $\tilde{\Delta F}_1^\text{el}$ has the same expression as in Eq.~\eqref{eq_AAII_dF1el}.
In the latter expression of the interaction energy, $v_\text{ii}$ plays the role of an effective ion-ion interaction potential and $v_\text{ie}$ plays the role of an effective ion-free-electron interaction potential. Provided that $Z>\sum_\alpha f_\alpha$, the three interaction potentials have a Coulomb tail. 

Still using the approximation of Eq.~\eqref{eq_one_center_approx}, the neutrality condition can be rewritten as:
\begin{align}
Z=\frac{N_\text{f}}{N_\text{i}}+\int_V d^3r\left\{n(\vec{r})\right\}
\label{eq_neutr_one_center}
\end{align}

The free energy per ion of our interacting-many-particle system can be written as:
\begin{align}
&F\left\{N_\text{f},\underline{w},\left\{f_\alpha\right\},\underline{v}_\text{trial};N_\text{i},V,T\right\}
\nonumber\\
&=\frac{1}{N_\text{i}}\int_V 
\frac{d^3R_1 d^3K_1...d^3r_{N_\text{f}} d^3k_{N_\text{f}}}{N_\text{i}!(2\pi)^{3N_\text{i}}N_\text{f}!(2\pi)^{3N_\text{f}}}
\left\{\vphantom{\frac{1}{2}}
\right.\nonumber\\
&w\left(N_\text{f},\vec{R}_1...\vec{K}_{N_\text{i}};\vec{r}_1...\vec{k}_{N_\text{f}}\right)
\nonumber\\
&\times\left(
\sum_{i=1}^{N_\text{i}} \frac{K_i^2}{2m_\text{nuc}} + 
\sum_{i=1}^{N_\text{f}} \frac{k_i^2}{2}
+U_\text{el}+N_\text{i}(\Delta F_1^0 +\Delta F_1^\text{xc})\right.
\nonumber\\
&+\left.\left.k_\text{B} T\ln w\left(N_\text{f},\vec{R}_1...\vec{K}_{N_\text{i}};\vec{r}_1...\vec{k}_{N_\text{f}}\right)
\vphantom{\sum_{i=1}^{N_\text{i}}}\right)\right\}
\label{eq_free_energy_before_thermo}
\end{align}
where $m_\text{nuc}$ is the nuclear-mass to electron-mass ratio, and $w\left(N_\text{f},\vec{R}_1...\vec{K}_{N_\text{i}};\vec{r}_1...\vec{k}_{N_\text{f}}\right)$ is the probability density of the classical state $(\vec{R}_1...\vec{K}_{N_\text{i}};\vec{r}_1...\vec{k}_{N_\text{f}})$ of the $N_\text{i}$ ions, with $N_\text{f}$ free electrons. 

In principle, the equilibrium state of the interacting-many-particle system would be found by minimizing $F$ with respect to $N_\text{f}$, $w$, $\{f_\alpha\}$, $v_\text{trial}$, while requiring the neutrality of Eq.~\eqref{eq_neutr_one_center} and the normalization of the probability density $w$.

In Eq.~\eqref{eq_free_energy_before_thermo}, the terms related to the shell structure can be separated from the terms related to the two-component classical fluid. In the latter terms, we can separate the ideal-gas terms from the excess free energy, we then get:
\begin{align}
F&=F_0^\text{i}(\underline{w};N_\text{i},V,T)+F_0^\text{e}(N_\text{f},\underline{w};N_\text{i},V,T)\nonumber\\
&+\Delta F_\text{ex}\left\{N_\text{f},\underline{w},\underline{n}\left\{\{f_\alpha\},\underline{v}_\text{trial};N_\text{i},V,T\right\}\right\}\nonumber\\
&+\Delta F_1^0
+\Delta F_1^\text{el}
+\Delta F_1^\text{xc}
\label{eq_AAII_DH_free_energy_before_th_limit}
\end{align}
The $\Delta F_\text{ex}$ term corresponds to the excess free energy per ion of the two-component classical plasma of ions and free electrons, with the interaction potentials $v_\text{ii}$, $v_\text{ie}$, $v_\text{ee}$, and no external potential (homogeneous system).

We consider the interacting-many-particle system in the thermodynamic limit, that is: $N_\text{i}\rightarrow\infty$, $N_\text{f}\rightarrow\infty$, $V\rightarrow\infty$ and $N_\text{i}/V=n_\text{i}$, $N_\text{f}/V=n_0$. The neutrality condition of Eq.~\eqref{eq_neutr_one_center} is then equivalent to Eq.~\eqref{eq_neutr_AAII}. In addition, the ion-free-electron potential becomes $v_\text{ie}\left\{\underline{n};r\right\}=v_\text{el}\left\{\underline{n};r\right\}$.

In the thermodynamic limit, a number of approximate theories exists, which allow one to address the case of a homogeneous classical fluid with arbitrary interaction potentials. Among these, we are particularly interested in those theories that replace the minimization of the free-energy of Eq.~\eqref{eq_AAII_DH_free_energy_before_th_limit} by another variational formulation, which consists in minimizing a free-energy functional of the pair correlation functions ${h}_\text{ii}$, ${h}_\text{ie}$, and ${h}_\text{ee}$. Using such a formulation, we are then left with the minimization of a free energy per ion with respect to ${h}_\text{ii}$, ${h}_\text{ie}$, ${h}_\text{ee}$, $\left\{f_\alpha\right\}$, ${v}_\text{trial}$, and $n_0$:
\begin{align}
F&\left\{\underline{h}_\text{ii},\underline{h}_\text{ie},\underline{h}_\text{ee},\left\{f_\alpha\right\},\underline{v}_\text{trial},n_0;n_\text{i},T\right\}\nonumber\\
&=F_0^\text{i}(n_\text{i},T)+F_0^\text{e}(n_0;n_\text{i},T)+\Delta F_1\left\{\left\{f_\alpha\right\},\underline{v}_\text{trial}\right\}\nonumber\\
&+\Delta F_\text{ex}\left\{\underline{h}_\text{ii},\underline{h}_\text{ie},\underline{h}_\text{ee},
\{f_\alpha\},\underline{v}_\text{trial},n_0;n_\text{i},T\right\}
\label{eq_AAII_DH_free_energy}
\end{align}

A free-energy functional such as $\Delta F_\text{ex}$, for a two-component classical fluid, was proposed for the hypernetted chain (HNC) approach in \cite{Lado73b,Enciso87}. However, since we are dealing with a two-component classical plasma, that is, a two-component classical fluid of particles having opposite charges, the HNC model faces the problem of the ``classical Coulomb catastrophe''. In such a physical context, only the Debye-H\"{u}ckel (DH) model allows one to circumvent the problem without resorting to a regularization of the potential not stemming from the model. However, this is done at the price of a linearization, which is always unjustified near the nucleus and which also limits strongly the validity domain of the approach. 

Free-energy functionals for the DH model of classical fluids were recently studied \cite{Piron16, Blenski17, Piron19a}. Using the most recent formula of \cite{Piron19a}, for the case of a two-component fluid, we have:
\begin{align}
&\Delta F_\text{ex}^\text{DH}
\left\{\underline{h}_\text{ii},\underline{h}_\text{ie},\underline{h}_\text{ee},
\{f_\alpha\},\underline{v}_\text{trial},n_0;n_\text{i},T\right\}
\nonumber\\
&=\frac{1}{n_\text{i}}A\left\{\bar{\bar{\underline{h}}}\{\underline{h}_\text{ii},\underline{h}_\text{ie},\underline{h}_\text{ee}, n_\text{i},n_0\},
\right.\nonumber\\
&\hspace{1.5cm}\left.\bar{\bar{\underline{u}}}\left\{\underline{n}\{\{f_\alpha\},\underline{v}_\text{trial}\},n_\text{i},n_0\right\};T
\vphantom{\bar{\bar{\underline{h}}}}\right\}\\
&A\left\{\bar{\bar{\underline{h}}},\bar{\bar{\underline{u}}};T\right\}
\nonumber\\
&=\frac{1}{2\beta}\int \frac{d^3k}{(2\pi)^3}\left\{
\text{Tr}\left(\bar{\bar{h}}_k\right)
-\ln\left(\text{det}\left(\mathbb{I}+\bar{\bar{h}}_k\right)\right)
+\beta \bar{\bar{h}}_k:\bar{\bar{u}}_k
\right\}
\end{align}
with the matrices of functions:
\begin{align}
&\bar{\bar{h}}_k\{\underline{h}_\text{ii},\underline{h}_\text{ie},\underline{h}_\text{ee}, n_\text{i},n_0\}\equiv
\left[
\begin{array}{c c}
  n_\text{i} h_{\text{ii},k} & \sqrt{n_\text{i}n_0} h_{\text{ie},k}  \\
  \sqrt{n_\text{i}n_0} h_{\text{ie},k} & n_0 h_{\text{ee},k}
\end{array}
\right]\\
&\bar{\bar{u}}_k\{\underline{n},n_\text{i},n_0\}
\equiv
\left[
\begin{array}{c c}
  n_\text{i} v_{\text{ii},k}\left\{\underline{n}\right\} & \sqrt{n_\text{i}n_0} v_{\text{ie},k}\left\{\underline{n}\right\}  \\
  \sqrt{n_\text{i}n_0} v_{\text{ie},k}\left\{\underline{n}\right\} & n_0 v_{\text{ee},k}
\end{array}
\right]
\end{align}
and where the ``$:$'' symbol denotes the Frobenius inner product of matrices, that is: $\bar{\bar{a}}:\bar{\bar{b}}=\sum_{i,j}a_{i,j}b_{i,j}$. $\beta\equiv 1/(k_\text{B}T)$ is the inverse temperature, and the Fourier transform $f_k$ of a function $f(r)$ is defined as follows:
\begin{align}
f_k\equiv\int d^3r\left\{f(r)e^{i\vec{k}.\vec{r}}\right\} 
\end{align}

From these expressions we can calculate the derivatives of $\Delta F_\text{ex}$ that are useful to the minimization procedure:
\begin{align}
\frac{\partial \Delta F_\text{ex}^\text{DH}}{\partial f_\alpha}
&=\frac{1}{n_\text{i}}\int d^3r d^3r'\left\{
\frac{\partial n(r)}{\partial f_\alpha}
\frac{{\delta \bar{\bar{u}}(r')}}{\delta n(r)}:
\frac{\delta A}{\delta \bar{\bar{u}}(r')}\right\}
\\
\frac{\delta \Delta F_\text{ex}^\text{DH}}{\delta v_\text{trial}(r)}
&=\frac{1}{n_\text{i}}\int d^3r' d^3r''\left\{
\frac{\delta n(r')}{\delta v_\text{trial}(r)}
\frac{{\delta \bar{\bar{u}}(r'')}}{\delta n(r')}:
\frac{\delta A}{\delta \bar{\bar{u}}(r'')}\right\}
\\
\frac{\partial \Delta F_\text{ex}^\text{DH}}{\partial n_0}
&=\frac{1}{n_\text{i}}\int d^3r\left\{
\frac{{\partial \bar{\bar{u}}(r)}}{\partial n_0}:
\frac{\delta A}{\delta \bar{\bar{u}}(r)}
+\frac{{\partial \bar{\bar{h}}(r)}}{\partial n_0}:
\frac{\delta A}{\delta \bar{\bar{h}}(r)}\right\}
\\
\frac{\delta \Delta F_\text{ex}^\text{DH}}{\delta h_{xx}(r)}
&=\frac{1}{n_\text{i}}\int d^3r' \left\{
\frac{{\delta \bar{\bar{h}}(r')}}{\delta h_{xx}(r)}:
\frac{\delta A}{\delta \bar{\bar{h}}(r')}\right\}
\end{align}
where $h_{xx}$ stands either for  $h_\text{ii}$, $h_\text{ie}$ or $h_\text{ee}$.

The free-energy functional $A$, is such that we have (see \cite{Piron19a}):
\begin{align}
&\frac{\delta A}{\delta \bar{\bar{h}}_k}=0\nonumber\\
&\Leftrightarrow
\left\{
\begin{array}{l}
h_{\text{ii},k} + n_\text{i} h_{\text{ii},k}\, \beta v_{\text{ii},k} + n_0 h_{\text{ie},k}\, \beta v_{\text{ie},k}
=-\beta v_{\text{ii},k}
\\
h_{\text{ee},k} + n_\text{i} h_{\text{ie},k}\,\beta v_{\text{ie},k} + n_0 h_{\text{ee},k}\,\beta v_{\text{ee},k}
=-\beta v_{\text{ee},k}
\\
h_{\text{ie},k} + n_\text{i} h_{\text{ii},k}\,\beta v_{\text{ie},k} + n_0 h_{\text{ie},k}\,\beta v_{\text{ee},k}
=-\beta v_{\text{ie},k}
\end{array}
\right.\label{eq_DH_eq_set}
\end{align}
that is: the two-component DH equation set, with the interaction potentials $v_\text{ii}$, $v_\text{ie}$, $v_\text{ee}$.

Due to the choice of free-energy functional (see the discussion on the non-uniqueness of the free-energy functional in \cite{Piron19a}, and Eq.~(55) therein), we have:
\begin{align}
\frac{\delta A}{\delta \bar{\bar{u}}(r)}=\frac{1}{2}\bar{\bar{h}}(r)
\end{align}
the latter form greatly simplifies the derivation of the model equations.

For the derivative of the interaction-potential matrix $\bar{\bar{u}}$, we have:
\begin{align}
\frac{{\delta \bar{\bar{u}}(r')}}{\delta n(r)}
=\left[
\begin{array}{c c}
  2n_\text{i}v_\text{el}(|\vec{r}'-\vec{r}|) 
  & \frac{\sqrt{n_\text{i}n_0}}{|\vec{r}'-\vec{r}|}   \\
   \frac{\sqrt{n_\text{i}n_0}}{|\vec{r}'-\vec{r}|}
  & 0
\end{array}
\right]
\end{align}
where $v_\text{el}$ is defined as in Eq.~\eqref{eq_def_vel}.

As in the previous section, in order to perform the minimization of the free energy $F$ while requiring the neutrality of Eq.~\eqref{eq_neutr_AAII}, we define the functional $\Omega$:
\begin{align}
\Omega&\left\{\underline{h}_\text{ii},\underline{h}_\text{ie},\underline{h}_\text{ee},\left\{f_\alpha\right\},\underline{v}_\text{trial},n_0;n_\text{i},T\right\}
\nonumber\\
&\equiv F+\mu\left(Z-\sum_\alpha f_\alpha-\frac{n_0}{n_\text{i}}\right)
\end{align}

Finally, performing the unconstrained minimization of $\Omega$:
\begin{align}
\frac{\delta\Omega}{\delta h_{\text{ii},k}}
=\frac{\delta\Omega}{\delta h_{\text{ie},k}}
=\frac{\delta\Omega}{\delta h_{\text{ee},k}}
=\frac{\delta\Omega}{\delta v_\text{trial}(r)}
=\frac{\partial\Omega}{\partial f_\alpha}
=\frac{\partial\Omega}{\partial n_0}=0
\end{align}
we get the equations:
\begin{align}
v_\text{trial}(r)=&v_\text{el}(r)+v_\text{xc}\left(n(r)\right)\nonumber\\
&+n_\text{i}\int d^3r' \left\{h_\text{ii}(r')v_\text{el}(|\vec{r}-\vec{r}'|)\right\}\nonumber\\
&+n_0\int d^3r' \left\{\frac{h_\text{ie}(r')}{|\vec{r}-\vec{r}'|}\right\}\label{eq_AAII_DH_vtrial}
\\
f_\alpha =& \frac{1}{e^{\beta(\varepsilon_\alpha-\mu)}+1}\label{eq_AAII_DH_FermiDirac}\\
\mu=&\mu_0(n_0,T)\nonumber\\
&+\frac{1}{2}\int d^3r\left\{ n_\text{i}h_\text{ie}(r)v_\text{ie}(r)+n_0h_\text{ee}(r)v_\text{ee}(r) \right\}\label{eq_AAII_DH_mu}
\end{align}
in addition to the DH equation set of Eq.~\eqref{eq_DH_eq_set}.

Equations \eqref{eq_AAII_DH_vtrial}, \eqref{eq_AAII_DH_FermiDirac}, \eqref{eq_AAII_DH_mu} differ from those of the isolated ion model, Eqs.~\eqref{eq_AAII_vtrial},\eqref{eq_AAII_FermiDirac},\eqref{eq_AAII_mu}, in two ways. First, the chemical potential of bound electrons $\mu$ now includes a correlation correction. This correction is simply the transposition of the correlation correction to the chemical potential of the free electrons, as it stems from the DH model.

Second, the self-consistent potential $v_\text{trial}$ now includes a contribution from the non-central ions (term involving $h_\text{ii}$) and free electrons (term involving $h_\text{ie}$). The latter correction have a strong impact on the self-consistent potential $v_\text{trial}$, since it leads to the screening of the potential. Rewriting Eq.~\eqref{eq_AAII_DH_vtrial} in the Fourier space, one gets:
\begin{align}
v_{\text{trial},k}=\left(1+n_\text{i}h_{\text{ii},k}\right)v_{\text{el},k}
+\frac{4\pi e^2}{k^2}n_0h_{\text{ie},k}
+v_{\text{xc},k}
\end{align}
Then, using the third equation of Eqs~\eqref{eq_DH_eq_set}, and the equalities $v_{\text{ie},k}=v_{\text{el},k}$, $v_{\text{ee},k}=4\pi/k^2$, we obtain:
\begin{align}
v_{\text{trial},k}&=\left(1+n_\text{i}h_{\text{ii},k}\right)v_{\text{el},k}
\frac{k^2}{k^2+4\pi\beta n_0}
+v_{\text{xc},k}\\
&=\left(1+n_\text{i}h_{\text{ii},k}\right)\left(Z-n_k\right)
\frac{4\pi}{k^2+4\pi\beta n_0}
+v_{\text{xc},k}
\end{align}
From the latter equation, we clearly see that there is no Coulomb term that remains in $v_{\text{trial}}$ when $r\rightarrow\infty$.

\section{Point-like ions and the continuum-lowering picture \label{sec_DHAAM_CL}}
Making the approximation of point-like ions amounts to replacing the interaction potentials $v_\text{ii}$ and $v_\text{ie}$ by their asymptotic forms, namely:
\begin{align}
v_\text{ii}\left\{\{f_\alpha\};R\right\}&\approx\frac{e^2}{R}\left(Z^*\right)^2\label{eq_approx_vii}
\\
v_\text{ie}\left\{\{f_\alpha\};r\right\}&\approx\frac{-e^2}{r}Z^*\label{eq_approx_vie}
\end{align}
with $Z^*$ being a shorthand notation for $Z^*\{f_\alpha\}\equiv Z-\sum_\alpha f_\alpha$.

Performing the minimization of $\Omega$ within this approximation, we obtain the following equations for the model:
\begin{align}
v_\text{trial}(r)=&v_\text{el}(r)+v_\text{xc}\left(n(r)\right)
\\
f_\alpha =& \frac{1}{e^{\beta(\varepsilon_\alpha+\Delta\varepsilon-\mu)}+1}\\
\Delta\varepsilon =& \int d^3r\left\{ \frac{e^2}{r} \left( n_0 h_\text{ie}(r) - n_\text{i} Z^* h_\text{ii}(r) \right) \right\}\\
\mu=&\mu_0(n_0,T)\nonumber\\
&+\frac{1}{2}\int d^3r\left\{\frac{e^2}{r}\left( n_0h_\text{ee}(r)-Z^*n_\text{i}h_\text{ie}(r) \right)\right\}
\end{align}
in addition to the DH equation set of Eq.~\eqref{eq_DH_eq_set}, with the pure Coulomb interaction potentials $v_\text{ii}$ and $v_\text{ie}$ of Eqs.~\eqref{eq_approx_vii}, \eqref{eq_approx_vie}.

Using the analytical solution of the two-component DH equations in the case of pure Coulomb interactions, namely:
\begin{align}
&h_\text{ii}(r)=-\beta e^2 (Z^*)^2 \frac{e^{-k_\text{D} r}}{r}\ ;\ h_\text{ie}(r)=+\beta e^2 Z^* \frac{e^{-k_\text{D} r}}{r}\nonumber\\
&h_\text{ee}(r)=-\beta e^2 \frac{e^{-k_\text{D} r}}{r}\ ;\ 
k_\text{D}\equiv\sqrt{4\pi\beta e^2(n_\text{i}Z^{*\,2}+n_0)}
\end{align}
we recover the results: 
\begin{align}
\Delta\varepsilon=&Z^*e^2 k_\text{D}\\
\mu=&\mu_0(n_0,T)-\frac{e^2k_\text{D}}{2}
\end{align}
which are the usual DH corrections to the ionization potential and chemical potential (see, for instance, \cite{Rouse62b}). However, we stress that the presence of the $\Delta\varepsilon$ correction to the orbital energies does not imply the suppression of bound states. 

Let us mention that several more sophisticated approaches to the continuum lowering were developed over the years (see \cite{EckerWeizel56,EckerKroll63,StewartPyatt66}), improving over this simple DH correction.




\section{One-component classical plasma correction \label{sec_DHAAM_OCP}}

\begin{figure}[t]
\centerline{\includegraphics[width=8cm]{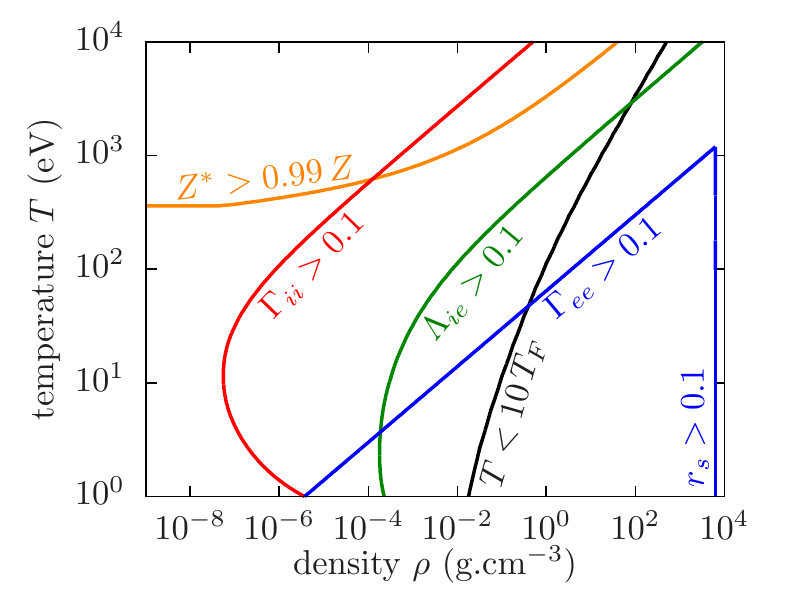}}
\caption{Indicative boundaries of significant ion-ion coupling ($\Gamma_\text{ii}=0.1$), electron-electron coupling (classical $\Gamma_\text{ee}=0.1$ and degenerate $r_\text{s}=0.1$), ion-electron polarization (electron-ion plasma parameter $\Lambda_\text{ie}=0.1$), electron degeneracy ($T=10\,T_\text{F}$), and partial ionization ($Z^*=0.99\,Z$). Mean ionization was taken from the Thomas-Fermi-Dirac model.
\label{fig_rho_T}}
\end{figure}

As it can be seen, for instance, in Fig.~\ref{fig_rho_T}, the low ion coupling is usually the most stringent validity condition of the ideal plasma approximation. This is especially true for high-Z elements.

In first approximation, one can then consider accounting only for the interaction between ions, keeping the ideal-gas approximation for the free electrons. This yields the picture of a one-component classical plasma (OCP) and $\Delta F_\text{ex}$  then is a OCP correlation contribution to the free-energy per ion. Since the OCP does not lead to the classical Coulomb catastrophe, this free energy can be taken to be either the DH or the HNC excess free energy.

In this approximation, the model equations become:
\begin{align}
v_\text{trial}(r)=&v_\text{el}(r)+v_\text{xc}\left(n(r)\right)\nonumber\\
&+n_\text{i}\int d^3r' \left\{h_\text{ii}(r')v_\text{el}(|\vec{r}-\vec{r}'|)\right\}
\\
f_\alpha =& \frac{1}{e^{\beta(\varepsilon_\alpha-\mu)}+1}\\
\mu=&\mu_0(n_0,T)
\end{align}
in addition to either the DH OCP equation:
\begin{align}
h_{\text{ii},k}=-\beta v_{\text{ii},k}-n_\text{i} h_{\text{ii},k}\, \beta v_{\text{ii},k}
\end{align}
or the HNC OCP equation set (Ornstein-Zernike and closure relations):
\begin{align}
&h_{\text{ii},k}=c_{\text{ii},k}+n_\text{i} h_{\text{ii},k}\, c_{\text{ii},k}\\
&c_{\text{ii}}(r)=-\beta v_{\text{ii}}(r)+h_{\text{ii}}(r)-\ln\left(h_{\text{ii}}(r)+1\right)
\end{align}
In the latter case, $c_{\text{ii}}$ is the ion-ion direct correlation function.

\begin{figure}[h]
\centerline{\includegraphics[width=8cm]{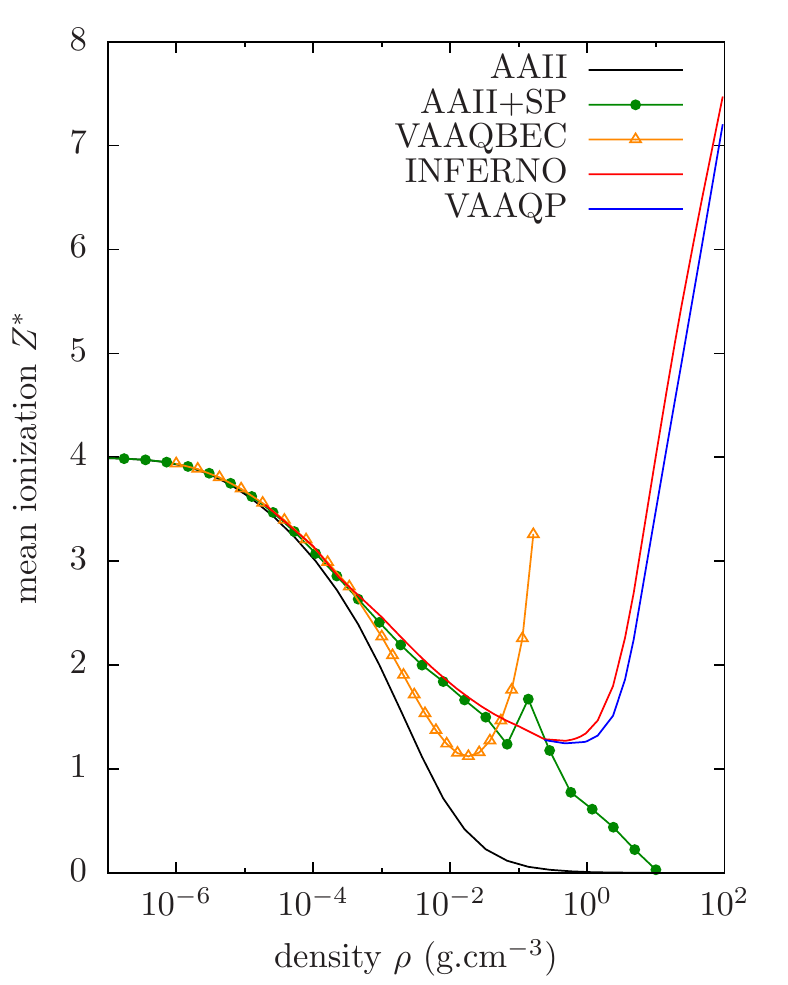}}
\caption{Mean ionization of silicon at 5 eV temperature, as calculated from the average-atom model of isolated ion (AAII), from the AAII with the Stewart-Pyatt continuum lowering and suppression of bound states (AAII+SP), from the VAAQBEC approach, from the INFERNO approach, and from the VAAQP approach.
\label{fig_mean_ioniz}}
\end{figure}

\section{Application and comparison with dense-plasma average-atom models\label{sec_Results}}

Fig.~\ref{fig_mean_ioniz} shows a comparison of mean ionizations obtained using various models, in the case of a warm, 5eV-temperature silicon plasma. The models are the average-atom model of isolated-ion (AAII, see Sec.~\ref{sec_AAII}), the same model with the heuristical Stewart-Pyatt continuum-lowering correction \cite{StewartPyatt66} (AAII-SP), the INFERNO model \cite{Liberman79,Liberman82} and the VAAQP model \cite{Blenski07b,Piron11}. In the latter two models, the mean ionization is defined from the electron chemical potential as follows:
\begin{align}
Z^*=\frac{\sqrt{2}}{n_\text{i}\pi^2\beta^{3/2}}\text{I}_{1/2}\left(\beta\mu\right)
\label{eq_def_zbar_mu}
\end{align}
where $\text{I}_{1/2}$ is the Fermi-Dirac integral of order $1/2$. This corresponds to the average number of electron per ion in the jellium.

As it can be seen in the figure, the AAII model exhibits the typical Saha-like behavior of decreasing mean ionization when density increases. 

The VAAQBEC approach of Sec.~\ref{sec_DHAAM} yields results that remain close to those of the VAAQP and INFERNO approaches up to densities of the order of $10^{-3}$ g.cm$^{-3}$. Beyond these densities, the decrease in mean ionization is more pronounced than in VAAQP or INFERNO. At even higher densities, the pressure ionization phenomenon occurs. It also occurs at lower densities in VAAQBEC than in VAAQP or INFERNO. 
This can be qualitatively related to the lack of free-electron degeneracy in this model, which may allow a stronger screening by free electrons. It can also be related to the impact of the DH approximation of the ion-ion correlation function, which may yield an overestimated probability of ions entering each other's Wigner-Seitz sphere.



\begin{figure}[t]
\centerline{\includegraphics[width=8cm]{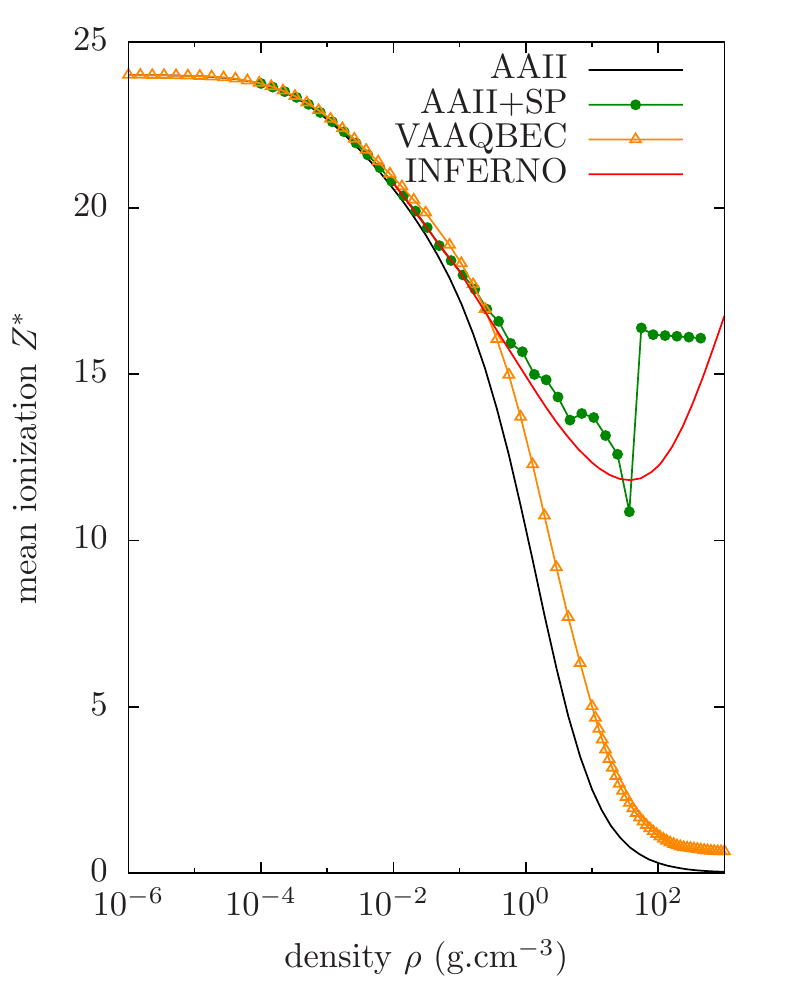}}
\caption{Mean ionization of iron at 200 eV temperature, as calculated from the average-atom model of isolated ion (AAII), from the AAII with the Stewart-Pyatt continuum lowering and suppression of bound states (AAII+SP), from the VAAQBEC approach, and from the INFERNO approach.
\label{fig_mean_ioniz_Fe}}
\end{figure}

Stewart-Pyatt continuum lowering, with suppression of bound states that lay above the continuum limit, allows the AAII model to mimic the results from the VAAQP and INFERNO models up to moderate densities, of the order of $10^{-2}$ g.cm$^{-3}$. Beyond such densities, strong pressure ionization occurs, the mean ionization departs from that of VAAQP and INFERNO, and the suppression of bound states can even lead to problems of convergence. In such regime, resonances in the continuum can be significantly populated. The mean ionization defined from Eq.~\eqref{eq_def_zbar_mu} differs significantly from $Z-\sum_\alpha f_\alpha$. The quantum treatment of continuum-electron polarization, which, to some extend, is accounted for in INFERNO and VAAQP models, is then required.

At even higher densities, of the order of $10^{-1}$ g.cm$^{-3}$, differences appear on the mean ionization between the INFERNO and VAAQP models. Such differences were discussed, for example, in \cite{Piron09c,Piron11,Piron11b}.

Fig.~\ref{fig_mean_ioniz_Fe} shows a comparison of mean ionizations in the case of a hot, 200eV-temperature iron plasma. The overall picture is quite similar to Fig.~\ref{fig_mean_ioniz}, except that, in this case, density effects occur at higher density, and that the VAAQBEC model does not seem to predict pressure ionization. The VAAQBEC approach yield results close to that of INFERNO up to densities of the order of 0.1 g.cm$^{-3}$. The Stewart-Pyatt continuum lowering give a mean ionization close to that of INFERNO up to densities of the order of 1 g.cm$^{-3}$.

Let us remark that the conditions of the experiments \cite{Bailey15} are: temperature close to 200 eV (180 eV estimated temperature), and free-electron densities about $3\cdot 10^{22}$ cm$^{-3}$. In Fig.~\ref{fig_mean_ioniz_Fe}, these conditions correspond to matter density around 0.15 g.cm$^{-3}$. These conditions then appear to be beyond the limit of applicability of the present model.


\section{Conclusion}
In this paper, we derive a variational average-atom model of electron-ion plasma with quantum treatment of bound electrons and accounting for correlations (VAAQBEC). This model allows one to address ion-ion, ion-electron and electron-electron correlations in a plasma, while determining self-consistently the ion average shell structure. Polarization of continuum electrons around ions is treated classically, through a two-component classical fluid approach.  

When ions are approximated by point-like particles, the present approach yields the usual Debye-H\"{u}ckel corrections to the orbital energies (continuum lowering) and chemical potential. If one considers only ion interactions and disregards the interactions of continuum electrons, the present approach yields ion-ion correlation corrections through a self-consistent one-component classical plasma contribution. 

Comparisons are presented with both the broadly-used continuum-lowering approach of Stewart and Pyatt \cite{StewartPyatt66} and with the dense-plasma average-atom models INFERNO \cite{Liberman79,Liberman82} and VAAQP \cite{Blenski07b,Piron11}. Results on warm (5 eV temperature) silicon, and hot (200 eV temperature) iron plasmas show agreement on the mean ionization with VAAQP and INFERNO in the low-coupling regime.
In this regime, the present model should in principle benefit from more realistic assumption on correlation functions. At higher coupling, the classical approach to correlations used in this model reaches the limits of its validity domain, which leads to a wrong estimation of the density effects on the shell structure.

Work is now in progress to include the effect of quantum polarization of continuum electrons.


\bibliographystyle{unsrt}

\end{document}